\documentstyle[12pt,epsfig]{article}

\begin{document}

\vspace{0.2cm}

\begin{flushright}
{\tt MPP-2012-084}
\end{flushright}

\vspace{0.2cm}

\begin{center}
{\large \bf Lepton Flavor Mixing Pattern and Neutrino Mass Matrix
\\ after the Daya Bay Experiment}
\end{center}

\vspace{0.2cm}
\begin{center}
{\bf Shun Zhou} \footnote{E-mail: zhoush@mppmu.mpg.de} \\
{\sl Max-Planck-Institut f\"{u}r Physik
(Werner-Heisenberg-Institut), D-80805 M\"{u}nchen, Germany}
\end{center}

\vspace{2.0cm}

\begin{abstract}
The Daya Bay Collaboration has recently observed neutrino
oscillations in the $\overline{\nu}^{}_e \to \overline{\nu}_e$
disappearance channel, indicating that $\sin^2 \theta^{}_{13} =
0.024\pm 0.005$ (1$\sigma$) and $\theta^{}_{13} = 0$ is already
excluded at the $5.2\sigma$ confidence level. Now three neutrino
mixing angles have been measured to a good degree of accuracy
($\theta^{}_{12} \approx 34^\circ$, $\theta^{}_{23} \approx
45^\circ$ and $\theta^{}_{13} \approx 9^\circ$). Motivated by these
experimental results, we propose a novel lepton flavor mixing
pattern, which predicts $\sin^2 \theta^{}_{23} = 1/2$, $\sin^2
\theta^{}_{12} = (2+\sqrt{3})/(10+\sqrt{3}) \approx 0.318$ and
$\sin^2 \theta^{}_{13} = (2-\sqrt{3})/12 \approx 0.022$, together
with a maximal CP-violating phase $\delta = 90^\circ$. The leptonic
CP violation characterized by the Jarlskog invariant ${\cal J} =
\sqrt{6}/72 \approx 3.4\%$ is promising to be measured in the future
long-baseline neutrino oscillation experiments. Furthermore, we
point out that a generalized version of $\mu$-$\tau$ symmetry may
exist in the neutrino sector and can give rise to the aforementioned
mixing pattern. The possible realizations in the seesaw models with
discrete flavor symmetries are also discussed.
\end{abstract}

\begin{center}
\end{center}

\newpage

\section{Introduction}

Recent years have seen dramatic progress in neutrino physics
\cite{XZ}. Now we have been convinced by a number of elegant solar,
atmospheric, accelerator and reactor neutrino experiments that
neutrinos are massive and lepton flavors are mixed \cite{PDG}. The
lepton flavor mixing can be described by a $3\times 3$ unitary
matrix $V$, i.e., the Maki-Nakagawa-Sakata-Pontecorvo (MNSP) matrix
\cite{MNSP}, which is conventionally parametrized through three
mixing angles ($\theta^{}_{12}$, $\theta^{}_{23}$,
$\theta^{}_{13}$), one Dirac-type CP-violating phase $\delta$ and
two Majorana-type CP-violating phases ($\rho$, $\sigma$). In the
standard parametrization \cite{PDG}, the MNSP matrix reads
\begin{eqnarray}
V = \left(\matrix{c^{}_{12} c^{}_{13} & s^{}_{12} c^{}_{13} &
s^{}_{13}e^{-i\delta} \cr -s^{}_{12} c^{}_{23} - c^{}_{12} s^{}_{23}
s^{}_{13} e^{i\delta} & c^{}_{12} c^{}_{23} - s^{}_{12} s^{}_{23}
s^{}_{13} e^{i\delta} & s^{}_{23} c^{}_{13} \cr s^{}_{12} s^{}_{23}
- c^{}_{12} c^{}_{23} s^{}_{13} e^{i\delta} & -c^{}_{12} s^{}_{23} -
s^{}_{12} c^{}_{23} s^{}_{13} e^{i\delta} & c^{}_{23}
c^{}_{13}}\right) \cdot P^{}_\nu \; ,
%     (1)
\end{eqnarray}
where $s^{}_{ij} \equiv \sin \theta^{}_{ij}$, $c^{}_{ij} \equiv \cos
\theta^{}_{ij}$ and $P^{}_\nu \equiv {\rm Diag}\{e^{i\rho},
e^{i\sigma}, 1\}$. If neutrinos are Dirac particles, the phase
matrix $P^{}_\nu$ can be transformed away by redefining the phases
of neutrino fields. From current neutrino oscillation data, one can
extract the experimentally favored ranges of mixing angles
($\theta^{}_{12}$, $\theta^{}_{23}$, $\theta^{}_{13}$), as well as
the neutrino mass-squared differences $\Delta m^2_{21} \equiv m^2_2
- m^2_1$ and $\Delta m^2_{31} \equiv m^2_3 - m^2_1$, where $m^{}_i$
(for $i = 1, 2, 3$) are neutrino masses. Unfortunately, there are so
far no constraints on the CP-violating phases. The absolute scale of
neutrino masses will be determined or constrained by the future
neutrinoless double beta decay experiments and cosmological
observations.

One decade ago, it was recognized that the lepton flavor mixing
might take the form of the so-called tri-bimaximal mixing pattern
(TBM) \cite{TB}:
\begin{eqnarray}
V^{}_0 = \left(\begin{array}{ccc} \displaystyle \frac{2}{\sqrt{6}} &
\displaystyle \frac{1}{\sqrt{3}} & 0 \\ \displaystyle -
\frac{1}{\sqrt{6}} & \displaystyle \frac{1}{\sqrt{3}} &
\displaystyle \frac{1}{\sqrt{2}} \\
-\displaystyle \frac{1}{\sqrt{6}} & \displaystyle \frac{1}{\sqrt{3}}
&- \displaystyle \frac{1}{\sqrt{2}}
\end{array}
\right) \; ,
%     (2)
\end{eqnarray}
implying $\sin^2 \theta^{}_{12} = 1/3$, $\sin^2 \theta^{}_{23} =
1/2$, $\sin^2 \theta^{}_{13} = 0$, and vanishing CP-violating
phases. Since the TBM pattern is well compatible with all the
neutrino oscillation experiments, it has stimulated a torrent of
model-building works based on the finite discrete flavor symmetries
\cite{group}.

In the middle of last year, the T2K \cite{T2K} and MINOS
\cite{MINOS} Collaborations released their data on neutrino
oscillation in the $\nu^{}_\mu \to \nu^{}_e$ appearance channel and
found a weak hint for a nonzero $\theta^{}_{13}$ with a significance
about $2\sigma$. The latest global-fit analysis of neutrino
oscillation experiments, including the T2K and MINOS data, yields
\cite{fit}
\begin{eqnarray}
\sin^2 \theta^{}_{12} = 0.312^{+0.017}_{-0.015} \; , ~~~~ \sin^2
\theta^{}_{23} = 0.52^{+0.06}_{-0.07} \; , ~~~~ \sin^2
\theta^{}_{13} = 0.013^{+0.007}_{-0.005} \; ,
%     (3)
\end{eqnarray}
in the case of normal neutrino mass hierarchy with $\Delta m^2_{21}
= [7.59^{+0.20}_{-0.18}] \times 10^{-5}~{\rm eV}^2$ and $\Delta
m^2_{31} = [2.50^{+0.09}_{-0.16}] \times 10^{-3}~{\rm eV}^2$ at the
$1\sigma$ confidence level. For the first time, the allowed range of
the CP-violating phase is given as $\delta = [-0.61^{+0.75}_{-0.65}]
\pi$. In the case of inverted neutrino mass hierarchy, the $1\sigma$
ranges for neutrino mixing angles are
\begin{eqnarray}
\sin^2 \theta^{}_{12} = 0.312^{+0.017}_{-0.015} \; , ~~~~ \sin^2
\theta^{}_{23} = 0.52^{+0.06}_{-0.06} \; , ~~~~ \sin^2
\theta^{}_{13} = 0.016^{+0.008}_{-0.006} \; ,
%     (4)
\end{eqnarray}
for neutrino mass-squared differences $\Delta m^2_{21} =
[7.59^{+0.20}_{-0.18}] \times 10^{-5}~{\rm eV}^2$ and $\Delta
m^2_{31} = -[2.40^{+0.08}_{-0.07}] \times 10^{-3}~{\rm eV}^2$, and
for the CP-violating phase $\delta = [-0.41^{+0.65}_{-0.70}] \pi$.
The fact that $\theta^{}_{13} > 0$ is obtained with a significance
about $3\sigma$ from the global analysis \cite{fit,Fogli} has led to
a large number of works that attempt to explain a relatively large
$\theta^{}_{13}$ \cite{aftT2K}.

Recently the Daya Bay Collaboration has announced the observation of
$\overline{\nu}^{}_e \to \overline{\nu}^{}_e$ disappearance and
found the evidence for a nonzero $\theta^{}_{13}$ with a $5.2\sigma$
significance \cite{Daya}. The best-fit value together with the
$1\sigma$ range is
\begin{equation}
\sin^2 \theta^{}_{13} = 0.024 \pm 0.005 \; ,
%      (5)
\end{equation}
which is consistent with the results from Double Chooz \cite{Chooz}
and RENO experiments \cite{Reno}, and the global-fit analysis as
well. The precise measurement of $\theta^{}_{13}$ is quite important
in the sense that (i) a relatively large $\theta^{}_{13}$ makes the
discovery of leptonic CP violation very promising in the future
long-baseline neutrino oscillation experiments, if the CP phase
itself is not highly suppressed; (ii) a relatively large
$\theta^{}_{13}$ indicates the significant deviation from the TBM
pattern, and may point to a different symmetry structure of lepton
flavor mixing \cite{aftDaya}.

Motivated by the new experimental results of $\theta^{}_{13}$, we
propose a novel neutrino mixing pattern as the alternative to the
TBM pattern:
\begin{eqnarray}
V^\prime = \left(\begin{array}{ccc} \displaystyle \frac{2}{\sqrt{6}}
& \displaystyle \frac{\sqrt{3} + 1}{2\sqrt{6}} & \displaystyle
-i\frac{\sqrt{3} - 1}{2\sqrt{6}} \\
\displaystyle - \frac{1}{\sqrt{6}} & \displaystyle
\frac{\sqrt{3}+1}{2\sqrt{6}}  - i\frac{\sqrt{3}-1}{4} &
\displaystyle \frac{\sqrt{3}+1}{4} -
i\frac{\sqrt{3}-1}{2\sqrt{6}} \\
-\displaystyle \frac{1}{\sqrt{6}} & \displaystyle
\frac{\sqrt{3}+1}{2\sqrt{6}} + i\frac{\sqrt{3}-1}{4} & \displaystyle
-\frac{\sqrt{3}+1}{4} - i\frac{\sqrt{3}-1}{2\sqrt{6}} \end{array}
\right) \; .
%     (6)
\end{eqnarray}
Comparing between $V^\prime$ and the standard parametrization in Eq.
(1), one can immediately obtain the maximal atmospheric mixing angle
$\sin^2 \theta^{}_{23} = 1/2$, a maximal CP-violating phase $\delta
= 90^\circ$, and
\begin{equation}
\sin^2 \theta^{}_{12} = \frac{2+\sqrt{3}}{10+\sqrt{3}} \approx 0.318
\; , ~~~~ \sin^2 \theta^{}_{13} = \frac{2-\sqrt{3}}{12} \approx
0.022 \; .
%     (7)
\end{equation}
It is straightforward to observe that the predictions for
$\theta^{}_{12}$ and $\theta^{}_{23}$ are very close to their
global-fit values in Eqs. (3) and (4), while the prediction for
$\theta^{}_{13}$ is in perfect agreement with the Daya Bay result in
Eq. (5). The Jarlskog invariant, which measures the magnitude of
leptonic CP violation, turns out to be ${\cal J} = \sqrt{6}/72
\approx 3.4\%$. The next important step in neutrino oscillation
experiments is to probe leptonic CP violation at the percent level.

The remaining part of this paper is organized as follows. In Sec. 2,
we reconstruct the neutrino mass matrix $M^{}_\nu$ by assuming the
mixing pattern in Eq. (6), and point out a generalized $\mu$-$\tau$
symmetry in the neutrino sector. In Sec. 3, we furthermore
illustrate how to obtain such a mixing pattern in a class of seesaw
models. Finally, we summarize our conclusions in Sec. 4.

\section{Generalized $\mu$-$\tau$ Symmetry}

The lepton flavor mixing arises from the mismatch in the
diagonalizations of charged-lepton and neutrino mass matrices. Once
the mixing matrix is determined from the neutrino oscillation
experiments, we can reconstruct the lepton mass matrices and learn
something about the underlying symmetry structure in the lepton
sector. However, it is obvious that such a reconstruction is not
unique and depends on the flavor basis we have chosen. In the basis
where the flavor eigenstates of charged leptons coincide with their
mass eigenstates, the neutrino mass matrix can be reconstructed from
neutrino mass eigenvalues and the MNSP matrix. Taking the TBM
pattern for example, we have
\begin{eqnarray}
M^0_\nu = V^{}_0 \left(\matrix{m^{}_1 & 0 & 0 \cr
                                            0 & m^{}_2 & 0 \cr
                                            0 & 0 & m^{}_3}\right)
                                            V^{\rm T}_0
\equiv \left(\matrix{a & b & b \cr
                b & c & d \cr
                b & d & c \cr}\right) \;,
%     (8)
\end{eqnarray}
where the four real parameters $(a, b, c, d)$ are linear
combinations of neutrino masses and satisfy the sum rule $a + b = c
+ d$. It is easy to verify that there exists a $\mu$-$\tau$ exchange
symmetry, which can be defined as $\nu^{}_{e{\rm L}}
\rightleftharpoons  \nu^{}_{e{\rm L}}$ and $\nu^{}_{\mu {\rm L}}
\rightleftharpoons  \nu^{}_{\tau {\rm L}}$ such that the neutrino
mass term is invariant under these transformations \cite{mt}. The
$\mu$-$\tau$ symmetry is responsible for the bimaximal mixing with
$\sin^2 \theta^{}_{23} = 1/2$ and $\theta^{}_{13} = 0$, while an
extra $Z^{}_2$ symmetry is needed to guarantee the condition $a + b
= c + d$, leading to the trimaximal mixing with $\sin^2
\theta^{}_{12} = 1/3$ \cite{Lam}. Since the charged-lepton mass
matrix in the chosen basis is diagonal, it should in general
preserve a $U(1)^3$ symmetry. Various finite discrete symmetry
groups, such as $A^{}_4$ and $S^{}_4$, have been invoked to derive
the TBM pattern at the leading order \cite{group,Lam}.

Now that a significant deviation from the TBM pattern has been
experimentally confirmed, we have to modify the TBM pattern, but in
the most economical way. The basis idea is that the overall neutrino
mass matrix can be decomposed into a symmetry-limit term and a
perturbation term, and it is the latter that induces corrections to
the TBM pattern and thus a nonvanishing $\theta^{}_{13}$. To account
for both a nonzero $\theta^{}_{13}$ and a maximal CP-violating
phase, we take the corrections to $V^{}_0$ as a rotation in the
$2$-$3$ complex plane:
\begin{eqnarray}
V = V^{}_0 \cdot \left(\matrix{1 & 0 & 0 \cr
                         0 & c^{}_\vartheta & -is^{}_\vartheta \cr
                         0 & -i s^{}_\vartheta & c^{}_\vartheta}\right)
= \left(\begin{array}{ccc} \displaystyle \frac{2}{\sqrt{6}} &
\displaystyle \frac{1}{\sqrt{3}} c^{}_\vartheta & \displaystyle
-\frac{i}{\sqrt{3}} s^{}_\vartheta \\
\displaystyle - \frac{1}{\sqrt{6}} & \displaystyle
\frac{1}{\sqrt{3}} c^{}_\vartheta - \frac{i}{\sqrt{2}}
s^{}_\vartheta & \displaystyle \frac{1}{\sqrt{2}} c^{}_\vartheta -
\frac{i}{\sqrt{3}} s^{}_\vartheta \\
-\displaystyle \frac{1}{\sqrt{6}} & \displaystyle \frac{1}{\sqrt{3}}
c^{}_\vartheta + \frac{i}{\sqrt{2}} s^{}_\vartheta & \displaystyle
-\frac{1}{\sqrt{2}} c^{}_\vartheta - \frac{i}{\sqrt{3}}
s^{}_\vartheta
\end{array} \right) \; ,
%     (9)
\end{eqnarray}
where $s^{}_\vartheta \equiv \sin \vartheta$ and $c^{}_\vartheta
\equiv \cos \vartheta$ with $\vartheta$ being a small rotation
angle. If $\vartheta = 0$ is taken, then the TBM in Eq. (2) is
reproduced. The above mixing pattern was first considered in Ref.
\cite{XZ07} in the framework of a minimal type-I seesaw model
\cite{mini}, and later discussed in Ref. \cite{Xing12,HZ} in light
of the experimental evidence for a nonzero $\theta^{}_{13}$. Another
simple but interesting case is to replace the rotation in $2$-$3$
plane with the one in $1$-$3$ plane, and its phenomenological
implications have been studied in detail in Ref. \cite{HS,FL}.

Compared with the standard parametrization in Eq. (1), such a mixing
pattern predicts a maximal atmospheric mixing angle $\theta^{}_{23}
= 45^\circ$ and an intriguing correlation between the other two
mixing angles \cite{XZ07}:
\begin{equation}
\sin^2 \theta^{}_{12} = \frac{1}{3} \left(1 - 2\tan^2
\theta^{}_{13}\right) \; .
%     (10)
\end{equation}
The salient feature of the relation in Eq. (10) is that
$\theta^{}_{12} \to 34^\circ$ as $\theta^{}_{13} \to 9^\circ$, which
are almost equal to their current best-fit values. Note that Eq.
(10) has also been obtained in Ref. \cite{smirnov} from the
constraints on the mixing matrix set by the structure of flavor
symmetry group. As an explicit example, we set $\vartheta = \pi/12$
in Eq. (9) and then get the mixing pattern $V^\prime$ in Eq. (6).
Its predictions have already been given in Eq. (7) and are in
excellent agreement with current oscillation data. And the simple
structure of $V^\prime$ is suggestive of some flavor symmetry.

Now we explore the implications of the MNSP matrix in Eq. (9) on the
symmetry structure of neutrino mass matrix. Given the mixing matrix
$V$ and the neutrino masses $m^{}_i$, one can immediately
reconstruct the neutrino mass matrix:
\begin{eqnarray}
M^{}_\nu = \frac{m^{}_1}{6} \left(\matrix{4 & -2 & -2 \cr -2 & 1 & 1
\cr -2 & 1 & 1}\right) + \frac{m^{}_2}{3} \left(\matrix{1 & 1 & 1\cr
1 & 1 & 1 \cr 1 & 1 & 1}\right) + \frac{m^{}_3}{2} \left(\matrix{0 &
0 & 0 \cr 0 & 1 & -1 \cr 0 & -1 & 1}\right) + \Delta M^{}_\nu \; ,
%     (11)
\end{eqnarray}
where $\Delta M^{}_\nu$ denotes the difference between the neutrino
mass matrix $M^{}_\nu$ and the counterpart $M^0_\nu$ in the TBM
case. More explicitly, we have
\begin{eqnarray}
\Delta M^{}_\nu = -\frac{m^{}_2+m^{}_3}{6} \sin^2 \vartheta
\left(\matrix{2 & 2 & 2 \cr
              2 & 5 & -1 \cr
              2 & -1 & 5}\right) -
i\frac{m^{}_2 + m^{}_3}{2\sqrt{6}} \sin 2\vartheta
\left(\matrix{0&1& -1 \cr
              1 & 2 & 0 \cr
              -1 & 0 & -2}\right) \; .
%     (12)
\end{eqnarray}
Note that $\Delta M^{}_\nu$ is proportional to $\sin \vartheta$, and
it will vanish when $\vartheta = 0$, as expected. One can also
easily figure out the neutrino mass matrix for the special case with
$\vartheta = \pi/12$, corresponding to the mixing matrix $V^\prime$
in Eq. (6). However, we shall focus on the general case as in Eq.
(11). It is straightforward to observe that the overall neutrino
mass matrix takes the following form
\begin{equation}
M^{}_\nu = \left(\matrix{x & y & y^* \cr
                  y & z & w \cr
                  y^* & w & z^* \cr}\right) \; ,
%     (13)
\end{equation}
where $(x, w)$ are real while $(y, z)$ are complex. In addition, two
constraint relations $x + {\rm Re}[y]  = w + {\rm Re}[z]$ and $2{\rm
Im}[y] = {\rm Im}[z]$ should be satisfied. The direct
diagonalization of $M^{}_\nu$ in Eq. (13) leads to
\begin{equation}
\tan 2\vartheta = - \frac{2\sqrt{6} {\rm Im}[y]}{{\rm Re}[y]+2{\rm
Re}[z]} \;.
%     (14)
\end{equation}
Once $\vartheta$ is fixed, the mixing angles are determined by
$\sin^2 \theta^{}_{13} = \sin^2 \vartheta/3$ and Eq. (10). The
remaining mixing parameters $\theta^{}_{23} = 45^\circ$ and $\delta
= 90^\circ$ arise from the intrinsic structure of $M^{}_\nu$ and
have nothing to do with the detailed values of the model parameters.
More general discussions about $M^{}_\nu$ in Eq. (13) without the
constraint conditions can be found in Ref. \cite{btc}.

We proceed to point out that a generalized $\mu$-$\tau$ symmetry
exists in the neutrino sector, which gives rise to the main features
of $M^{}_\nu$ in Eq. (13). In the chosen flavor basis, the neutrino
mass term can be explicitly written as
\begin{eqnarray}
{\cal L}^{}_\nu &=& -\frac{1}{2} \sum_{\alpha, \beta}
\left(M^{}_\nu\right)^{}_{\alpha \beta} \overline{\nu^{}_{\alpha
{\rm L}}} \nu^c_{\beta {\rm L}}+ {\rm h.c.} \; ,
%     (15)
\end{eqnarray}
where $\left(M^{}_\nu\right)^{}_{\alpha \beta}$ (for $\alpha, \beta
= e, \mu, \tau$) stand for the matrix elements of $M^{}_\nu$. If we
define the generalized $\mu$-$\tau$ symmetry by
\begin{eqnarray}
\nu^{}_{e{\rm L}} \rightleftharpoons  \nu^c_{e{\rm L}} \;, ~~~~~~
\nu^{}_{\mu{\rm L}} \rightleftharpoons  \nu^c_{\tau{\rm L}} \;,
~~~~~~ \nu^{}_{\tau{\rm L}} \rightleftharpoons  \nu^c_{\mu{\rm L}}
\;,
%     (16)
\end{eqnarray}
and require the neutrino mass term to be invariant under the above
transformations, then the matrix elements
$\left(M^{}_\nu\right)^{}_{\alpha \beta}$ are found to fulfill the
following conditions:
\begin{eqnarray}
\left(M^{}_\nu\right)^{}_{ee} = \left(M^{}_\nu\right)^*_{ee}, ~~~
\left(M^{}_\nu\right)^{}_{\mu \tau} = \left(M^{}_\nu\right)^*_{\mu
\tau}, ~~~ \left(M^{}_\nu\right)^{}_{e\mu} =
\left(M^{}_\nu\right)^*_{e\tau}, ~~~ \left(M^{}_\nu\right)^{}_{\mu
\mu} = \left(M^{}_\nu\right)^{*}_{\tau\tau}\; .
%     (17)
\end{eqnarray}
These conditions are satisfied exactly by the neutrino mass matrix
$M^{}_\nu$ in Eq. (13). However, extra flavor symmetries or
empirical assumptions have to be introduced to enforce the
constraint relations among the four independent matrix elements.

\section{Implications for Model Building}

In the previous discussions, we have put aside the mechanism for
neutrino mass generation. One of the simplest scenarios to
accommodate tiny neutrino masses is to extend the standard model by
three right-handed neutrinos $N^{}_{i{\rm R}}$ for $i = 1, 2, 3$,
which are singlets under the electroweak gauge group ${\rm
SU}(2)^{}_{\rm L} \times {\rm U}(1)^{}_{\rm Y}$. In this scenario,
the general lepton mass mass terms are
\begin{equation}
{\cal L}^{\rm I}_\ell = - \overline{l^{}_{\rm L}} M^{}_l E^{}_{\rm
R} - \overline{\nu^{}_{\rm L}} M^{}_{\rm D} N^{}_{\rm R} -
\frac{1}{2} \overline{N^c_{\rm R}} M^{}_{\rm R} N^{}_{\rm R} + {\rm
h.c.} \; ,
%     (18)
\end{equation}
where $M^{}_l$ is the charged-lepton mass matrix, $M^{}_{\rm D}$ and
$M^{}_{\rm R}$ are the Dirac and Majorana neutrino mass matrices,
respectively. Note that $M^{}_{\rm R}$ is the mass matrix for gauge
singlets $N^{}_{\rm R}$ and thus it is not subject to the
electroweak gauge symmetry breaking at the scale $\Lambda^{}_{\rm
EW} \sim 100~{\rm GeV}$. For ${\cal O}(M^{}_{\rm R}) \gg
\Lambda^{}_{\rm EW}$, the effective neutrino mass matrix is given by
$M^{}_\nu \approx - M^{}_{\rm D} M^{-1}_{\rm R} M^{\rm T}_{\rm D}$,
i.e., the type-I seesaw formula \cite{type1}. Therefore, the
smallness of light neutrino masses is ascribed to the largeness of
the masses of heavy Majorana neutrinos.

In the flavor basis where $M^{}_l$ and $M^{}_{\rm R}$ are diagonal,
i.e., $M^{}_l = {\rm Diag}\{m^{}_e, m^{}_\mu, m^{}_\tau\}$ and
$M^{}_{\rm R} = {\rm Diag}\{M^{}_1, M^{}_2, M^{}_3\}$, we apply the
generalized $\mu$-$\tau$ symmetry to the light neutrino fields
$\nu^{}_{\rm L}$ as in the previous section. Furthermore, we assume
the lepton mass terms are also invariant under $N^{}_{i{\rm R}}
\rightleftharpoons N^c_{i{\rm R}}$. In this case, the Dirac neutrino
mass matrix turns out to be
\begin{equation}
M^{}_{\rm D} = \left(\matrix{ a^{}_1 & a^{}_2 & a^{}_3 \cr b^{}_1
e^{+i\varphi^{}_1} & b^{}_2 e^{+i\varphi^{}_2} & b^{}_3
e^{+i\varphi^{}_3} \cr b^{}_1 e^{-i\varphi^{}_1} & b^{}_2
e^{-i\varphi^{}_2} & b^{}_3 e^{-i\varphi^{}_3}}\right) \; ,
%     (19)
\end{equation}
where $a^{}_i$, $b^{}_i$ and $\varphi^{}_i$ (for $i = 1, 2, 3$) are
real parameters. One can verify that the effective neutrino matrix
$M^{}_\nu$ via the seesaw formula takes exactly the expected form in
Eq. (13) with
\begin{equation}
x = \sum^3_{i = 1} \frac{a^2_i}{M^{}_i} \;, ~~~~ y = \sum^3_{i = 1}
\frac{a^{}_i b^{}_i }{M^{}_i} e^{i\varphi^{}_i} \;, ~~~~ w =
\sum^3_{i = 1} \frac{b^2_i}{M^{}_i} \;, ~~~~ z = \sum^3_{i = 1}
\frac{b^2_i}{M^{}_i} e^{2i\varphi^{}_i} \; .
%     (20)
\end{equation}
If $a^{}_i = b^{}_i \cos \varphi^{}_i$ holds for $i = 1, 2, 3$, the
two constraint conditions $x+{\rm Re}[y] = w+{\rm Re}[z]$ and ${\rm
Im}[z] = 2{\rm Im}[y]$ can be satisfied. Consequently, we obtain the
mixing matrix in Eq. (9).

As another example, we now turn to the type-II seesaw model, where a
${\rm SU}(2)^{}_{\rm L}$ triplet scalar $\Delta$ is introduced to
generate a Majorana mass term for three light neutrinos
\cite{type2}. In this model, the lepton mass terms are
\begin{equation}
{\cal L}^{\rm II}_\ell = - \overline{l^{}_{\rm L}} M^{}_l E^{}_{\rm
R} - \frac{1}{2} \overline{\nu^{}_{\rm L}} M^{}_\nu \nu^{c}_{\rm L}
+ {\rm h.c.} \; ,
%     (21)
\end{equation}
where $M^{}_\nu = Y^{}_\Delta \langle \Delta \rangle$ with
$Y^{}_\Delta $ of ${\cal O}(1)$ and $\langle \Delta \rangle \sim
0.1~{\rm eV}$ being the neutrino Yukawa coupling matrix and the
vacuum expectation value of the neutral component of the triplet
scalar, respectively. Therefore, the smallness of vacuum expectation
value $\langle \Delta \rangle$ is responsible for the tiny neutrino
masses.

In order to obtain the desired mixing pattern in Eq. (9), we follow
a phenomenological approach and assume that the charged-lepton
mass matrix $M^{}_l$ is real (i.e., $M^{}_l = M^*_l$), and it
furthermore satisfies two commutation relations:
\begin{equation}
\left[M^{}_l, {\cal D}\right] = {\bf 0}, ~~~~~~~~~ \left[M^{}_l,
{\cal P}^{}_{\mu \tau}\right] = {\bf 0} \; ,
%     (22)
\end{equation}
where the matrices ${\cal D}$ and ${\cal P}^{}_{\mu \tau}$ are defined as
\begin{equation}
{\cal D} = \frac{1}{3} \left(\matrix{1 & 1 & 1\cr
                         1 & 1 & 1\cr
                         1 & 1 & 1}\right) \; , ~~~~
{\cal P}^{}_{\mu \tau} = \left(\matrix{1 & 0 & 0\cr 0 & 0 & 1\cr 0 &
1 & 0}\right) \; .
%     (23)
\end{equation}
As observed in Ref. \cite{HS}, the charged-lepton mass matrix
$M^{}_l$ under these conditions can be exactly diagonalized by the
TBM pattern, i.e., $V^{\rm T}_0 M^{}_l V^{}_0 = {\rm Diag}\{m^{}_e,
m^{}_\mu, m^{}_\tau\}$. On the other hand, we impose a $Z^{}_2$
symmetry on the neutrino fields, under which $\nu^{}_{e{\rm L}}$ is
odd and $\nu^{}_{\alpha{\rm L}}$ (for $\alpha = \mu, \tau$) are
even. Or equivalently, the neutrino mass matrix $M^{}_\nu$ fulfills
the commutation relation $[M^{}_\nu, {\cal G}^{}_\nu] = {\bf 0}$ with
\begin{equation}
{\cal G}^{}_\nu = \left(\matrix{-1 & 0 & 0 \cr 0 & 1 & 0 \cr 0 & 0 &
1}\right) \; .
%     (24)
\end{equation}
Hence the neutrino mass matrix is block diagonal
\begin{equation}
M^{}_\nu = \left(\matrix{
                   A & 0 & 0 \cr
                   0 & B & C  \cr
                   0 & C & D}
\right) \; .
%     (25)
\end{equation}
Without loss of generality, one can make $A$, $B$ and $D$ real by
redefining the phases of neutrino fields. Such a redefinition will
contribute to the Majorana-type CP-violating phases of the MNSP
matrix, which are not of our interest here. Hence only the parameter
$C$ is complex, and the Dirac CP-violating phase is in general
arbitrary. If $C$ is assumed to be imaginary (i.e., $C = i
\tilde{C}$ with $\tilde{C}$ real), the neutrino matrix $M^{}_\nu$ in
Eq. (25) can be diagonalized as follows
\begin{equation}
\left(\matrix{1 & 0 & 0 \cr 0 & c^{}_\vartheta & -i s^{}_\vartheta
\cr 0 & -i s^{}_\vartheta & c^{}_\vartheta}\right)^\dagger
\left(\matrix{
                   A & 0 & 0 \cr
                   0 & B & i\tilde{C}  \cr
                   0 & i\tilde{C} & D}
\right) \left(\matrix{1 & 0 & 0 \cr 0 & c^{}_\vartheta & -i
s^{}_\vartheta \cr 0 & -i s^{}_\vartheta & c^{}_\vartheta}\right)^*
= \left(\matrix{ m^{}_1 & 0 & 0\cr 0 & m^{}_2 & 0 \cr 0 & 0 &
m^{}_3}\right) \; ,
%     (26)
\end{equation}
where the rotation angle is given by $\tan 2\vartheta =
-2\tilde{C}/(B+D)$, while the neutrino masses are
\begin{eqnarray}
m^{}_1 &=& A \; , \nonumber \\
m^{}_2 &=& \frac{1}{2} \left[\sqrt{(D+B)^2 + 4\tilde{C}^2} -
(D-B)\right]\; , \nonumber \\
m^{}_3 &=& \frac{1}{2} \left[\sqrt{(D+B)^2 + 4\tilde{C}^2} +
(D-B)\right] \; .
%     (27)
\end{eqnarray}
As we have already seen, the mixing pattern in Eq. (9) with
$\vartheta = \pi/12$ is well consistent with current neutrino
oscillation data. Therefore, the relation $B + D + 2\sqrt{3}
 \tilde{C} = 0$ holds in this case. Such a relation, together with
the absolute neutrino mass $m^{}_1$ and two independent neutrino
mass-squared differences $(\Delta m^2_{21}, \Delta m^2_{31})$,
uniquely determines the model parameters.

\section{Summary}

Motivated by the recent experimental evidence for a relatively large
$\theta^{}_{13}$, we suggest a novel neutrino mixing pattern that
predicts $\sin^2 \theta^{}_{12} = (2+\sqrt{3})/(10+\sqrt{3}) \approx
0.318$, $\sin^2 \theta^{}_{23} = 1/2$, $\sin^2 \theta^{}_{13} =
(2-\sqrt{3})/12 \approx 0.022$ and a maximal CP-violating phase
$\delta  = 90^\circ$. These predictions are very close to the
best-fit values of $\theta^{}_{12} \approx 34^\circ$,
$\theta^{}_{23} \approx 45^\circ$ and $\theta^{}_{13} \approx
9^\circ$. The leptonic CP violation, characterized by the Jarlskog
invariant ${\cal J} = \sqrt{6}/72 \approx 3.4\%$, is expected to be
measured in the future long-baseline neutrino oscillation
experiments. Such a mixing pattern can be viewed as the minimal
modification of the well-known tri-bimaximal mixing pattern through
the rotation in $2$-$3$ complex plane with a small rotation angle
$\vartheta = \pi/12$. If one identifies $\vartheta$ with the Cabibbo
angle $\theta^{}_{\rm C} \approx 13^\circ$, the resultant MNSP
matrix should also be compatible with current neutrino oscillation
experiments. In this case, it might be possible to relate the flavor
mixing in the lepton sector to that in the quark sector.

In the general case with an arbitrary angle $\vartheta$, we
reconstruct the neutrino mass matrix $M^{}_\nu$ in the flavor basis
where the charged-lepton mass matrix is diagonal. Furthermore, we
point out that there exists a generalized $\mu$-$\tau$ symmetry,
defined by the transformations $\nu^{}_{e{\rm L}} \rightleftharpoons
\nu^c_{e{\rm L}}$ and $\nu^{}_{\mu {\rm L}} \rightleftharpoons
\nu^c_{\tau {\rm L}}$, in the neutrino sector. Two examples in the
type-I and type-II seesaw models have been worked out to derive the
desired neutrino mass matrix in a phenomenological way. We believe
that this investigation should be helpful in understanding the
lepton flavor mixing with a relatively large $\theta^{}_{13}$. In
particular, the imposed symmetries in the charged-lepton and
neutrino sectors in these two examples may be instructive for the
model building based on some discrete flavor symmetries.

The ongoing and forthcoming neutrino oscillation experiments will
provide us with more precise values of three neutrino mixing angles,
so the predictions of the suggested mixing pattern are easily to be
confirmed or refuted by the future oscillation data. The
maximal CP violation at the percent level will be soon tested in the
long-baseline oscillation experiments. In any case, the experimental
hints on neutrino mixing parameters are definitely desirable and
will finally guide us to the true theory of fermion masses, flavor
mixing and CP violation.

\vspace{0.8cm}

\begin{flushleft}
{\large \bf Acknowledgements}
\end{flushleft}

{\sl The author would like to thank Zhi-zhong Xing for helpful
comments and discussions, and Georg Raffelt for warm hospitality at
the Max-Planck-Institut f\"{u}r Physik, M\"{u}nchen.}

\end{document}